# Role of Molecular Structure in Defining the Dynamical Landscape of Deep Eutectic Solvents at Nanoscale

T. Rinesh [1,2], H. Srinivasan[1,2], V.K. Sharma[1,2], V. García Sakai[3] and S. Mitra[1,2*]

[1]Solid State Physics Division, Bhabha Atomic Research Centre, Mumbai 400085, India

[2]Homi Bhabha National Institute, Anushaktinagar, Mumbai 400094, India

[3] ISIS Neutron and Muon Centre, Rutherford Appleton Laboratory, Didcot, UK

* Corresponding author: smitra@barc.gov.in

Tel:+91-22-25594674; FAX: +91-22-25505151

# ABSTRACT

The molecular dynamics of deep eutectic solvents (DESs) are complex, characterized by nanoscale spatial and temporal heterogeneity. Understanding these dynamics is crucial for tailoring transport properties like diffusion, viscosity and ionic conductivity. Molecular diffusion in DESs stems from transient caging and translation jumps, necessitating an understanding of how molecular structure regulates these processes. This study explores the influence of alkyl chain length on the nanoscopic dynamics of alkylamide-lithium perchlorate based DESs using quasielastic neutron scattering (QENS) and molecular dynamics (MD) simulations. QENS results show that, despite its shorter chain length and lighter mass, acetamide (ACM) exhibited the lowest mobility among the alkylamides, including propanamide (PRM) and butyramide (BUT). Detailed analysis of QENS data reveals that long-range jump diffusion is fastest in ACM and slowest in BUT, essentially due to their differences in molecular size, mass and also enhanced complexation in longer alkyl chain molecules. However, the localized dynamics follows an unusual trend, where PRM is the fastest and ACM is the slowest. Despite greater flexibility in BUT, the slower caged dynamics impedes its localized motion. These findings highlight the interplay between alkyl chain length and DES dynamics, emphasizing role of molecular structure in governing transport properties.

1. INTRODUCTION

Solvents are vital for a number of physical and chemical processes across numerous industries, significantly influencing efficiency and effectiveness.[1, 2] Traditional organic solvents suffer inherent toxicity and high volatility posing threats to human health and the environment,[3, 4] thus the development of sustainable and safer alternatives has become critical. Neoteric solvents such as supercritical carbon dioxide (s$CO_2$), liquid polymers, gas-expanded solvents (GXLs), and ionic liquids (ILs) have emerged as promising options.[3, 5] Among these, deep eutectic solvents (DESs) have recently gained significant attention.[3, 6]

DESs are recognized for their low toxicity,[7, 8] minimal environmental impact[7, 9] and straightforward synthesis. These solvents are produced by combining a salt with hydrogen bond donors (HBDs) in precise molar ratios, resulting in a substantial drop in the freezing point of the mixture.[10-12] At this composition, known as the eutectic point, extensive intermolecular hydrogen bonding between the components of the mixture.[10, 12] DESs offer a range of favorable physicochemical properties, including a wide liquidus range, low vapor pressure, water compatibility, suppressed flammability and high thermal stability. These characteristics make DESs highly suitable for diverse applications, such as metal extraction and electrodeposition, chemical and material synthesis, catalysis and various separation processes.[11-13] The physicochemical properties and stability of DESs are critically determined by their constituents. HBDs such as glycerols, alkylamides and urea have drawn considerable attention in this context.[6, 14, 15] Studies on aqueous mixture of DESs have also yielded promising results, as these solvents are nonreactive with water. The inclusion of water alters the intermolecular bonding, microscopic dynamics and the structural organization of the molecules, thereby impacting the macroscopic transport properties of the mixture such as viscosity and conductivity.[16-18]

Rising interest in electrolyte-based systems for electrochemical applications has driven advancements in high-voltage lithium-ion batteries utilizing $Li^+$ salt-based electrolytes, enhancing performance and overcoming key challenges.[19, 20] This progress has extended to DESs synthesized from amides and lithium salts showing promise as potential electrolytes.[21-23] Acetamide-based DESs, formulated with lithium salts such as lithium bis(trifluoromethanesulfonyl)imide (LiTFSI) and lithium perchlorate (Li$ClO_4$), exhibit high electrochemical stability, making them promising electrolytes for double-layer capacitors.[21] DESs made of alkylamides (RCON$H_2$, R=$CH_3$, $CH_3CH_2$, $CH_3CH_2CH_2$) and lithium salts (LiX, X=$ClO_4^-$, $NO_3^-$ and $Br^-$) at 78:22 molar composition form liquid solutions at room temperatures, but their practical applicability is constrained by their high viscosity and low ionic conductivity.[24] These limitations are primarily due to the strong interactions between alkylamide molecules and lithium ions, leading to the formation of stable ion-pair complexes.[25] In particular, studies on acetamide with lithium perchlorate have revealed that the diffusive properties of these DESs strongly correlate with the lifetime of the complexes.[18, 26] Specifically, the self-diffusion coefficient of alkylamides exhibits a power-law dependence on complexation lifetime.

Experimental and simulation studies on these systems have highlighted significant heterogeneity in the solution structure and dynamics.[24, 27] Structural analysis highlights the formation of clusters with a variety of sizes and lifetimes, showing a pronounced anionic dependence.[28] $Li^+$ ions, in combination with anions of salt, drive the structural heterogeneity in these systems, where a strong positive dependence on the alkylamide chain length is observed.[15] Furthermore, the dynamics within DESs are inherently complex, characterized by significant spatial and temporal heterogeneities.[18, 29] Understanding their dynamics at the molecular level is essential to correlate to the resulting transport properties, such as diffusion, viscosity, and ionic conductivity, which are critical for optimizing their performance. Notably, DESs containing lithium bromide ($LiBr^-$) exhibit greater heterogeneity compared to those with nitrate ($NO_3^-$) or perchlorate ($ClO_4^-$) as anions.[30] Time resolved fluorescence spectroscopic measurements,[24] dielectric relaxation experiments[31] and MD simulations[30] have further revealed a strong viscosity decoupling in these systems with weakest for DESs containing perchlorate ion and strongest for longer amide chains.[24] However, it is notable that there is a strong scale dependence in the viscosity-diffusion relationship, as recently discovered for these DESs.[30] This kind of scale-dependent diffusion-viscosity decoupling is also prevalent in ionic liquids.[32] The influence of anion identity is further evident in diffusion dynamics, where the self-diffusion coefficient of alkylamides are lowest in DES with bromide ions and highest with perchlorate ions.[33] In parallel, solvation dynamics studies have shown a notable slowdown in the average solvation rate when nitrate ions are replaced with bromide ions. These findings highlight the role of anionic identity of the salt in shaping the overall dynamical behavior of DESs.[27]

Further, numerous studies have also focussed on the effect of molecular structure of the constituents in amide based DESs.[34-40] These studies reflect that altering the alkyl chain affects physical properties such as density, refractive index, thermal conductivity, melting point etc.[34-37] More importantly, increasing the chain length strengthens molecular interactions, as observed in both the alkyl chain length variation of alcohols[36] and the cations of salt.[34] This enhanced interaction leads to changes in macroscopic properties such as viscosity, surface tension, and interfacial structure.[37] Dynamically, longer alkyl chains are associated with a reduction in molecular mobility, reflected in slower solvation dynamics and lower diffusion coefficients.[38, 40] Furthermore, chain length modulates spatial and dynamical heterogeneity within DES, with increasing chain length resulting in greater structural heterogeneity.[39, 40] Thus, understanding the impact of chain length provides an additional degree of control over the physicochemical properties of DESs.

However, the influence of chain length in electrolyte-based DESs, particularly those composed of alkylamides and lithium salts, is not well explored. This study aims to bridge this gap by examining the effect of varying alkylamide chain lengths in DESs formed with lithium perchlorate. The choice of $ClO_4^-$ as the anionic component is motivated by its peculiar ability to break molecular structures in aqueous media.[41] Furthermore, lithium perchlorate, due to its high polarity, exhibits exceptional solubility in organic solvents and enhances the kinetic performance of DESs.[42]

In this study, we employ quasielastic neutron scattering (QENS) experiment in conjunction with atomistic MD simulations to study the effect of alkyl chain length on the dynamics of alkylamide based electrolytes. QENS is a powerful technique that directly probes the dynamics at length scales of a few Å and time scales ranging from picosecond to nanoseconds.[43] The simultaneous spatial and temporal sensitivity of QENS allows for a detailed understanding of the molecular-level dynamics. By integrating QENS with classical

MD simulations, which operate on similar length and time scales, we gain accurate atomistic insights into the underlying dynamics of these systems.

We examined the effect of alkyl chain length on the dynamics of three alkylamides: acetamide, propanamide and butyramide, (hereafter referred as ACM, PRM and BUT respectively) each mixed with lithium perchlorate salt at a molar ratio of 78:22. Given that the interaction between alkylamides and lithium ions significantly contributes to the elevated viscosity of DESs and limiting molecular mobility,[25] the nature of molecular complexation between alkylamides and lithium ions as a function of chain length was also analysed to understand how it evolves with chain length. Our study highlights the importance of molecular structure in modulating the underlying dynamics within DESs, revealing how variations in chain length influences the interactions and transport properties, which are critical to optimize performance.

## 2. Experimental Section

*Materials and Sample Preparation:* The synthesis of DESs was carried out by combining solid mixtures of alkylamides, namely acetamide, propanamide and butyramide with lithium perchlorate at a molar ratio of 78:22. The mixture was then heated to 340 K for approximately 3 hours to yield a clear and homogeneous solution, which was stable upon cooling to room temperature (300K).

*Quasielastic Neutron Scattering (QENS):* QENS experiments were conducted using the high-resolution, near backscattering inverted geometry spectrometer, IRIS at the ISIS Pulsed Neutron and Muon Source, Rutherford Appleton Laboratory, U.K.[44] The spectrometer provides an energy resolution of ~17.5μeV (full width at half maximum) using a pyrolytic graphite (002) analyzer crystal. The instrument was operated in offset mode, with an accessibly energy and wave-vector ($Q$) transfer ranges of -0.3 to +1.0 meV and 0.5 to 1.8 Å$^{-1}$, respectively. The measurements were done at 330 and 365 K respectively. The energy resolution of the instrument was determined by performing QENS measurements on a standard vanadium sample and data reduction was carried out using MANTID software.[45]

*Simulation Setup:* Atomistic MD simulations were performed using the NAMD simulation package[46] on a mixture of alkylamides (RCONH$_2$) and lithium perchlorate-based DESs. The alkylamides included acetamide (CH$_3$CONH$_2$), propanamide (CH$_3$CH$_2$CONH$_2$), and butyramide (CH$_3$CH$_2$CH$_2$CONH$_2$). The number of alkylamide and lithium perchlorate molecules were fixed at 400 and 56 respectively to maintain a molar ratio of 78:22. The initial configuration for each system was built using PACKMOL[47] by randomly arranging the molecules within a cubical simulation box. The interactions between the molecules were parameterized using the CHARMM forcefield[48] for acetamide, propanamide and butyramide. For the ions, force fields were taken from existing literature.[49, 50] All simulations were performed in NPT ensemble using Langevin thermostat[51] and barostat[52] under 1 atm pressure and a temperature of 365 K. For temperature and system density equilibration, the simulation setup was equilibrated for 5 ns. Following which the production run was executed for 20 ns, with trajectories recorded every 1 ps. Periodic boundary conditions were imposed in all three directions and the equations of motion were integrated using velocity-Verlet algorithm,[53] with a 1 fs time step. A separate simulation was also conducted to capture shorter trajectories at smaller recording frequencies of 0.1 and 0.01 ps for 1 ns and 100 ps respectively.

3. **Theoretical Background**

In neutron scattering experiments, the intensity of the scattered neutron is measured as a function of momentum transfer ($Q = k_f - k_i$) and energy transfer ($E$). This intensity is described by the double differential scattering cross section, which is composed of coherent and incoherent contributions, reflecting the interactions between neutrons and the sample

$$\frac{d^2\sigma}{dE\,d\Omega} \propto \frac{k_f}{k_i}[\sigma_{coh}\,S_{coh}(\boldsymbol{Q},E) + \sigma_{inc}\,S_{inc}(\boldsymbol{Q},E)] \quad (1)$$

where $\sigma_{coh}$ and $\sigma_{inc}$ are the coherent and incoherent scattering cross sections while $S_{coh}(\boldsymbol{Q},E)$ and $S_{inc}(\boldsymbol{Q},E)$ denote the corresponding scattering functions. Due to the large incoherent scattering cross section of hydrogen, the neutron scattering signal from hydrogen atoms dominates, effectively masking the contribution from other atoms in hydrogen-rich systems.[43] As a result, for such systems, eq (1) simplifies to:

$$\frac{d^2\sigma}{dE\,d\Omega} \propto \frac{k_f}{k_i}[\,\sigma_{inc}\,S_{inc}(\boldsymbol{Q},E)] \quad (2)$$

The incoherent scattering function, $S_{inc}(\boldsymbol{Q},E)$, provides insight into self-diffusion processes, reflecting both the spatial and temporal aspects of individual atomic motions. This scattering function is related to the van Hove self-correlation function, $G_s(\boldsymbol{r},t)$, through a Fourier transform in space and time[43] providing a connection between experimental observables and atomic-level dynamics.

$$S_{inc}(\boldsymbol{Q},E) = \int_{-\infty}^{\infty} dt\, e^{-iEt/\hbar} \int d^3\boldsymbol{r}\, e^{-i\boldsymbol{Q}\cdot\boldsymbol{r}}\,G_s(\boldsymbol{r},t) \quad (3)$$

For isotropic systems such as liquids, the incoherent scattering function, $S_{inc}(Q,E)$, is averaged over all possible orientations of $Q$, due to the isotropic nature of scattering. This averaging simplifies the analysis of neutron scattering data in liquid systems, making QENS a particularly useful technique for studying molecular diffusion and other dynamic processes. The combination of QENS spectra and MD simulations provides a powerful approach for understanding the dynamic behavior of systems, offering complementary insights within the same spatial and temporal scales.

4. **Results and Discussion**

Typical QENS spectra showing the effects of alkyl chain length on the microscopic dynamics for all three DESs (RCONH$_2$ + LiClO$_4$) at 330 K and at a representative $Q$ are shown in Figure 1. Notably, the QENS data in all these three systems captures the diffusion of alkylamide molecules, owing to high incoherent scattering cross section of hydrogen atoms as discussed previously. The energy resolution of the instrument which marks the time scale of the slowest motion is shown in dashed lines. The broadening of the elastic line, referred to as the quasielastic signal, serves as an indicator of the presence of stochastic dynamics observable within the resolution of the QENS spectrometer. The extent of the broadening in the quasielastic peak correlates directly with the molecular mobility in the system. As seen in Figure 1, significant dynamics are evident within the spatio-temporal regime of the IRIS spectrometer, as indicated by pronounced quasielastic broadening. The dynamical behavior presented in Figure 1 reveals an interesting observation. Despite ACM being the smallest molecule among the three, acetamide based DESs exhibits the least quasielastic broadening, indicating the slowest dynamics. In contrast, the propanamide-based DES, represented in blue, shows the greatest broadening, suggesting it has the fastest dynamics. To better understand this counterintuitive behavior, we study the QENS spectra by modeling it explicitly using a microscopic diffusion mechanism that involves contributions from different degrees of freedom.

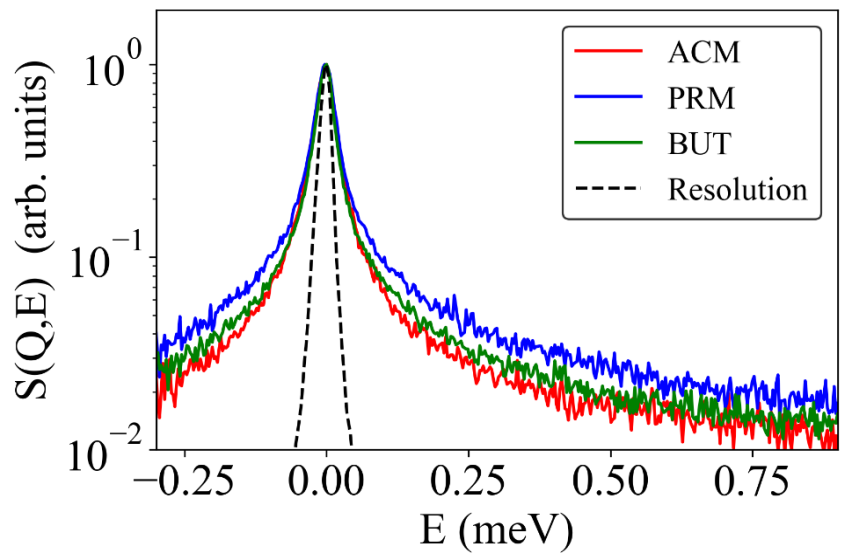

**Figure 1.** *QENS spectra of DES based on ACM (in red), PRM (in blue) and BUT (in green) shown at a Q-value of 1.2 Å$^{-1}$ and at 330 K. The instrument resolution measured using vanadium standard is shown in black dashed lines. The spectra shown here are peak normalized to establish a direct comparison of the broadening*

The molecular diffusion in DESs have been shown to follow cage-jump diffusion mechanism.[25, 30, 54] This mechanism is primarily attributed to the formation of transient cage-like structures, followed by jump diffusion process.[25, 30, 54] This diffusion behavior is a hallmark of liquids governed by extensive hydrogen bonding networks, where caging occurs as a result of hydrogen bond formation or complexation with neighboring molecules.[55, 56] Within these cages, the molecules exhibit localized diffusion, constrained by the interactions with their immediate environment. The extent of caging is largely dependent on the lifetime of these hydrogen bonds or complexes, as longer-lived interactions lead to more pronounced confinement. Upon reaching the average bond lifetime, the molecular cage relaxes, allowing the molecule to escape and undergo a jump to another cage and form hydrogen bonds with a new set of neighboring molecules. This cage-jump mechanism plays a crucial role in determining the system's transport dynamics, affecting both localized and long-range molecular mobility.

Based on this, a two-component diffusion model has been used to model the QENS spectra, effectively capturing the observed dynamical characteristics in these DESs. This model has proved successful across a broad range of complex systems[14, 57-59] and has been rigorously validated for ACM-based DES through previous studies using a similar approach that integrates MD[25] and QENS.[25, 30] The two motions are manifested as a convolution of two scattering laws: one associated with long-range jump diffusion of the alkylamide center of mass (COM) and the other incorporating the localized translation diffusion of hydrogen atoms within a spherical confinement. Furthermore, the latter can be reckoned as a combination of internal dynamical dynamics about the molecular COM and caged motion of the molecular COM. While internal dynamics arise from molecular rotations and chain flexibility, the caged dynamics of alkylamides are due to the long-lived complexes and hydrogen bonds formed in the DESs. Accordingly, the scattering law is expressed as:

$$S(Q,E) = L_j(\Gamma_j, E) \otimes \left[ C_0(Q)\delta(E) + (1 - C_0(Q)) L_{loc}(\Gamma_{loc}, E) \right] \quad (4)$$

Here, $L_j$ and $L_{loc}$ are Lorentzians pertaining to the jump and localized diffusion processes. The relaxation time scales associated to these motions can be inferred from their half-width half-maxima (HWHM) of $\Gamma_j$ and $\Gamma_{loc}$ respectively, whereas the information of the geometry of internal motions can be inferred from $C_0(Q)$ which represents the elastic incoherent structure factor (EISF). When convoluted with instrument resolution, $R(E)$, the scattering law used to describe the quasielastic data attains the form:

$$S(Q,E) = \left[ C_0(Q)L_j(\Gamma_j, E) + (1 - C_0(Q)) L_{j+loc}(\Gamma_{j+loc}, E) \right] \otimes R(E) \quad (5)$$

Typical fitted QENS spectra for RCONH$_2$ + LiClO$_4$ DES at 330 K and $Q$=1.2 Å$^{-1}$ along with the individual contributions from the components are shown in Figure 2. To gain quantitative insight into these motions, a detailed analysis of the $Q$-dependence of the fitted parameters, $C_0(Q)$, $\Gamma_j$ and $\Gamma_{loc}$ were conducted.

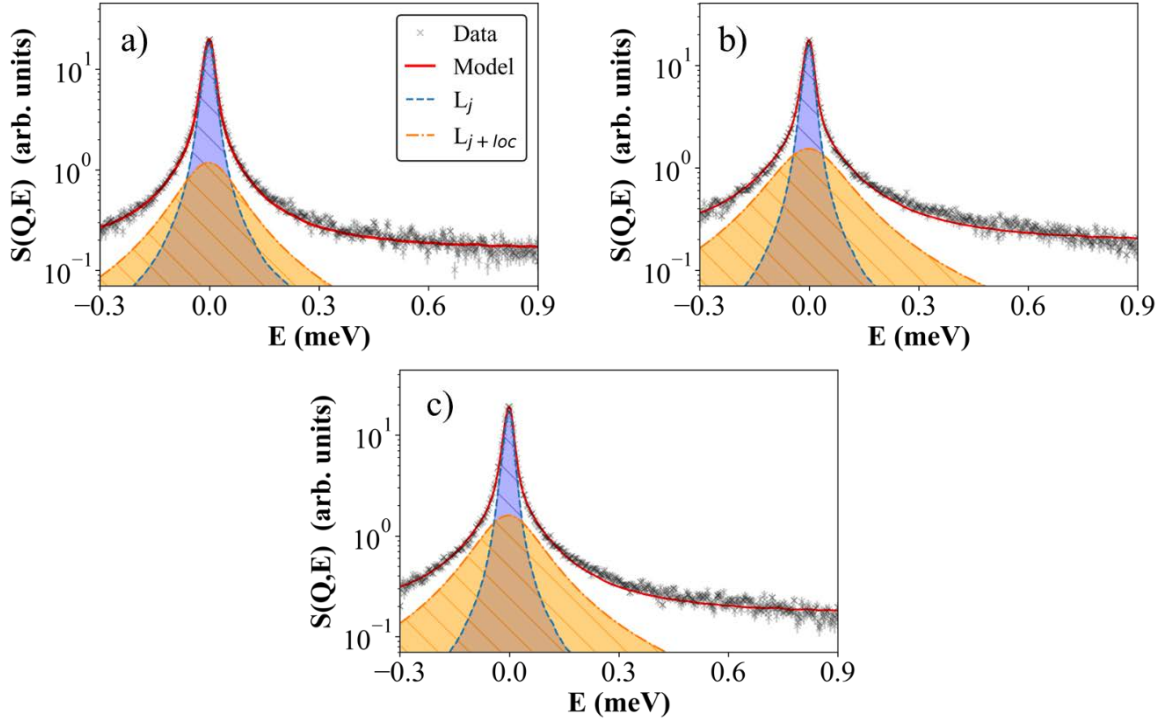

**Figure 2.** *QENS spectra of DES based on a) ACM b) PRM and c) BUT shown at a Q-value of 1.02 Å$^{-1}$ at 330 K. The individual components of the model are shown by shaded regions*

**4.1 Jump Diffusion:** A detailed understanding of the long-range jump diffusion process is crucial for characterizing mass transport in the system. Thus, the HWHMs of the Lorentzian corresponding to jump diffusion, $\Gamma_j$, for the COM motion of the alkylamide molecules were studied over the entire $Q$ range for all three systems at 330 and 365 K as shown in Figure 3. Since $\Gamma_j$ is directly related to the jump-diffusion process, it demonstrates that increasing alkyl chain length results in a deceleration of the system's jump dynamics. This is expected, since with increase in alkyl chain length the molecules tend to have greater inertia and therefore the jumps don't occur frequently.[60-62] The $Q^2$ dependence of $\Gamma_j(Q)$ can be effectively modelled using Singwi-Sjölander jump diffusion model.[63] This model describes the diffusion of molecules as a series of discrete jumps at random time intervals, with jump lengths ($l_0$) following an exponential distribution. For this case, the $Q$ dependence of $\Gamma_j$ is given by,

$$\Gamma_j(Q) = \frac{D_j Q^2}{1 + D_j Q^2 \tau_j} \quad (6)$$

with the jump diffusivity $D_j = l_0^2/6\tau_j$ and $\tau_j$ the average time interval between two consecutive jumps. In the limit where the jump frequency increases significantly i.e. as $\tau_j \to 0$, the particle begins to exhibit continuous Brownian motion. In this limit, $\Gamma_j(Q)$ asymptotically approaches $D_j Q^2$. This quadratic dependance of $\Gamma_j(Q)$ on $Q$ is a distinct signature of Gaussian diffusive behavior. This behavior is captured in the low $Q$ regime which reflects longer spatial displacements $\left(Q \sim \frac{1}{r}\right)$, where the number of jumps becomes large, and the particle's motion enters the Gaussian regime.[18] However, as $Q$ increases, reflecting shorter spatial scales, the influence of localized motions becomes significant, leading to a deviation from this quadratic dependence. This deviation marks the breakdown of Gaussian behavior, suggesting that at higher $Q$-values, the dynamics are no longer purely diffusive in nature. The least-squares fittings of $\Gamma_j$ using Eq. 6 for each of the DESs are also shown in Figure 3. The effectiveness of the model is demonstrated by its consistent fitting results across all three DESs at the measured temperatures. As shown in Table 1, acetamide exhibits the highest jump diffusion coefficient compared to other DES at the measured temperatures. However, this value is considerably smaller than that observed in its molten form.[25] A similar order in the diffusivity values of DES components has been observed both experimentally and through simulations for various choline chloride based DESs.[64-66] The set of dynamical parameters associated with the jump diffusion process as a function of alkyl chain length for all three DESs across the two measured temperatures is listed in Table 1. The jump diffusivities exhibit a decreasing trend with increasing alkyl chain length. This trend is consistent with observations in other systems, where the influence of alkyl chain length of the cation in the dynamics of teraalkylammonium chloride and decanoic acid based DESs was studied.[67] This behavior is in line with the slowing down of Stokes shift dynamics observed in alkylamide and lithium perchlorate DESs with longer alkyl chains, further supporting this behavior.[24]

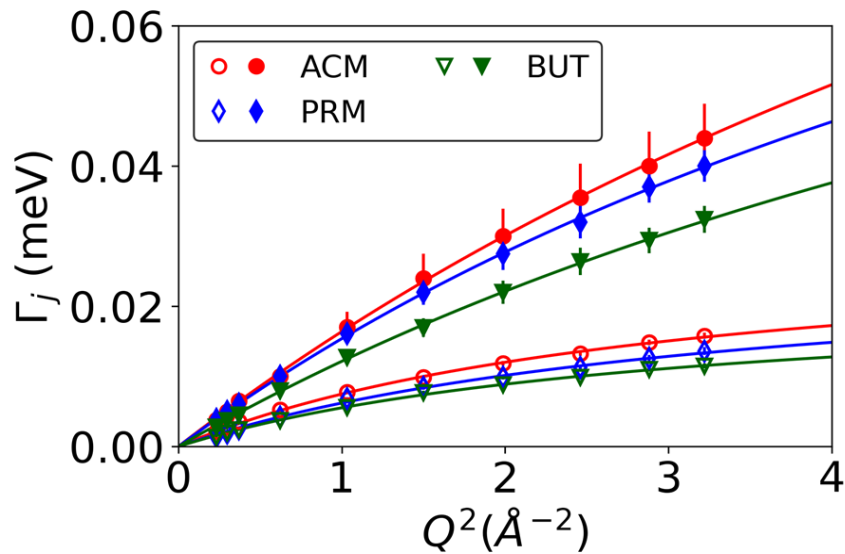

**Figure 3.** *Quasielastic widths ($\Gamma_j$) of the Lorentzian function ascribed to a jump diffusion process for all DESs, obtained from spectral fitting at 330 K (open symbols) and 365 K (filled symbols). The solid lines are the fits based on SS model of jump diffusion given in Eq. 6.*

**Table I.** List of Dynamical Parameters obtained from modelling the jump diffusion of alkylamides from QENS data at 330 and 365 K

| Alkylamide | 330 K | | 365 K | |
|---|---|---|---|---|
| | $D_j$ (x $10^{-6}$ cm$^2$/s) | $\tau_j$ (ps) | $D_j$ (x $10^{-6}$ cm$^2$/s) | $\tau_j$ (ps) |
| Acetamide | 1.50 ($\pm$ 0.04) | 21.53 ($\pm$ 0.72) | 2.74 ($\pm$ 0.05) | 3.61 ($\pm$ 0.24) |
| Propanamide | 1.21 ($\pm$ 0.05) | 23.69 ($\pm$ 1.56) | 2.62 ($\pm$ 0.04) | 4.68 ($\pm$ 0.23) |
| Butyramide | 1.11 ($\pm$ 0.04) | 28.99 ($\pm$ 1.26) | 2.04 ($\pm$ 0.04) | 5.26 ($\pm$ 0.34) |

### 4.1.1 Molecular complexation of Alkylamide in DES

The long-range jump diffusion processes in DES are significantly influenced by the molecular complexation between the constituents. In systems composed of mixtures of acetamide and lithium perchlorate salts, complexation of acetamide with lithium ions plays a critical role in slowing the jump dynamics of acetamide.[25] Interestingly, a marked enhancement in dynamics was observed when the lifetime of these complexes was modulated.[18, 26] Therefore, it becomes essential to systematically investigate how alkylamide-lithium ion complexation evolves with increasing alkyl chain length and whether it continuously regulates the jump diffusion dynamics of alkylamides. In this regard, MD simulations were conducted on a DES system of alkylamides and LiClO$_4$ at 365 K, examining the underlying bond dynamics between alkylamides and lithium ions.

Given the strong tendency of Li$^+$ to form complexes with the carbonyl oxygen in alkylamides,[25] the bond dynamics between the carbonyl oxygen of alkylamides and Li$^+$ were analyzed using the time-dependent bond autocorrelation function, $C(t)$. This function quantifies the probability that a bond present at an initial time $t_0$, remains intact at a later time $t$. The presence or absence of the bond is tracked using a flag operator, $h(t)$, which is assigned a value of 1 if a complex exists between the alkylamide and Li$^+$, or 0 otherwise. Mathematically, the bond autocorrelation function is expressed as:

$$C(t) = \sum_{pairs} \frac{\langle h(t+t_0)h(t_0) \rangle}{\langle h^2(t_0) \rangle} \qquad (7)$$

where the angular brackets denote averaging over different time origins $t_0$. The temporal evolution of $C(t)$ for different DESs is illustrated in Figure 4. As shown in the figure, the correlation function decays more slowly with increasing alkyl chain length, indicating that longer chain alkylamides form more stable complexes with Li$^+$. This slower decay suggests that the complexation between alkylamides and lithium ions becomes stronger as the alkyl chain length increases. This strengthening of alkylamide-Li$^+$ complexation with increasing chain length was further corroborated by the analysis of the intermolecular pair distribution function (PDF), $g_{Li^+-O_A}(r)$, between lithium ions and the oxygen sites of alkylamides. As shown in Figure S1, transitioning from acetamide to propanamide leads to an increase in the area under the PDF curve. This increase further confirms our observation, reflecting strengthening of interactions between lithium ions and alkylamide oxygen atoms with increasing alkyl chain length.

To further quantify these observations, the bond autocorrelation function was modelled using a stretched exponential function, $e^{\left(-\frac{t}{\tau}\right)^\beta}$ with $\tau$ and $\beta$ as fitting parameters.[68] The average lifetime of the complexes, $\tau_c$, was subsequently estimated from the fitted parameters using the relation $\tau_c = \frac{\tau}{\beta} \Gamma(\frac{1}{\beta})$. The estimated values of $\tau_c$ were 0.45 ns for ACM, 0.72 ns for PRM and 1.2 ns for BUT. Interestingly, with each increment in alkyl chain length, $\tau_c$ exhibits a consistent increase by a factor of approximately 3.5 from ACM to BUT. The smaller the value of $\tau_c$, the shorter-lived these complexes are. However, the progressively increasing $\tau_c$ values with longer alkyl chains indicate enhanced stability and strength of the alkylamide-Li$^+$ complexes as the chain length grows. Similar Li$^+$ ion complexes, typically lasting several nanoseconds, have been reported in ionic liquid-based electrolytes, where their lifetimes are influenced by the nature of the organic solvent and by salt concentration.[69, 70] Notably, temperature has an inverse impact on the complexation lifetimes, with longer lived complexes observed at higher temperatures.[69-71]

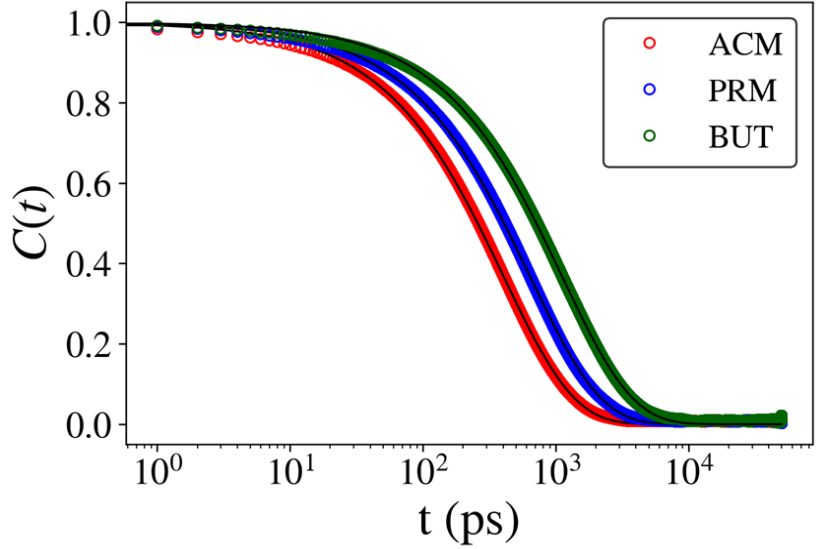

**Figure 4.** *Bond autocorrelation for alkylamide-Li$^+$ for different DESs, calculated from MD simulation trajectories using Eq. (7). The black solid lines are the individual fits*

Given that molecular complexation between alkylamides and lithium ions plays a critical role in governing jump dynamics, the reduced mobility in longer-chain alkylamides can be attributed not only to their increased molecular mass but also to the enhanced complexation strength. This stronger complexation significantly impedes their dynamics, contributing to the overall slowing of jump dynamics.

**4.2 Internal Motions:** In addition to the jump diffusion process, the internal dynamics of the molecules are observable in the time scales of QENS. These internal motions are parametrized by an EISF ($C_0(Q)$) and a HWHM ($\Gamma_{loc}$), both obtained from fitting the QENS data. The internal motions encompass localized diffusion of the entire molecule within transient cages, isotropic rotational diffusion, as well as various conformational changes, including bending and torsional mobility. The combination of these motions can be effectively modelled by assuming that each hydrogen atom undergoes a localized translation diffusion (LTD) within a soft spherical confinement. The scattering law for soft confinement is given as:[72]

$$S_{loc}^{LTD}(Q,E) = A_0(Q)\,\delta(E) + \frac{1}{\pi}\sum_{m=1}^{\infty} A_m(Q)\,\frac{(mD_{loc}/\sigma^2)}{(mD_{loc}/\sigma^2)^2 + E^2} \quad (8)$$

The elastic contribution is described by the first term on the right-hand side, whereas the quasielastic contributions, consisting of multiple Lorentzian functions, are accounted for by the second term. Here, $A_0(Q)$ and $A_m(Q)$ correspond to the elastic and quasielastic structure factors, respectively. $D_{loc}$ denotes the diffusion coefficient and $\sigma$ quantifies the size of the domain in which the particles are diffusing, the radius of the soft confining sphere. While,

Eq. (8) is the scattering law for LTD within a soft sphere of single radius, $\sigma$. In this system, owing to its known large heterogeneity[26, 30], we are assuming an exponential distribution of confinement spheres of different radii with an average value of $\sigma_0$. Under such assumptions, the model function for the EISF is given by

$$A_0(Q) = \int_0^\infty d\sigma \, \frac{e^{-(\sigma/\sigma_0)}}{\sigma_0} \, e^{-(Q\sigma)^2}$$

$$A_0(Q) = \frac{\sqrt{\pi}}{2Q\sigma_0} e^{\frac{1}{4Q^2\sigma_0^2}} erfc\left(\frac{1}{2Q\sigma_0}\right) \quad (9)$$

The EISF, $C_0(Q)$, obtained from the QENS spectra at 330 K and 365 K for the three DESs, are fitted using Eq. 9, with $\sigma_0$ as the fitting parameter. The corresponding model fit is shown in Figure 5(a) which is very satisfactory. Increasing the chain length of alkyl molecules alters the EISF, suggesting that the geometry of the localized diffusion in alkylamides is perturbed by chain length variation. Notably, the average cage size, $\sigma_0$, is largest for PRM, followed by BUT, with ACM exhibiting the smallest. These results are consistent for the two measured temperatures as given in Table II.

The dynamical behavior of localized motions can be understood by studying the variation of $\Gamma_{loc}$ as a function of $Q$. Modelling this variation provides quantitative insights into these dynamics. As evident in Figure 5(b), the slower localized dynamics in ACM can be attributed to its shorter chain length. Shorter chains result in reduced molecular flexibility, giving ACM fewer conformational degrees of freedom compared to its longer chain counterparts. With increasing chain length, an enhancement in molecular flexibility and a broader range of conformational freedom are expected, leading to faster localized motions in BUT than in PRM. However, this trend is not reflected in the QENS data when transitioning from PRM to BUT. This non-monotonic behavior of localized dynamics can be quantified, by fitting $\Gamma_{loc}$, with $\Gamma_{loc} \to D_{loc}/\sigma^2 = 1/\tau$ in the low Q limit, where $\tau$ characterizes the timescale of these motions. From Table II, we clearly note that the associated timescales follow the trend $\tau_{PRM} < \tau_{BUT} < \tau_{ACM}$, indicating fastest localized motions in PRM and slowest in ACM. This unusual behavior could stem from the fact that the observed localized dynamics within the instrument window reflect contributions not only from the molecular flexibility but also from the diffusion of molecules within transient cages. Previous studies have noted that both these dynamical features occur at similar timescales [25, 54] making them indistinguishable in QENS experiments. In an attempt to investigate the individual contributions, we employ atomistic level MD simulations.

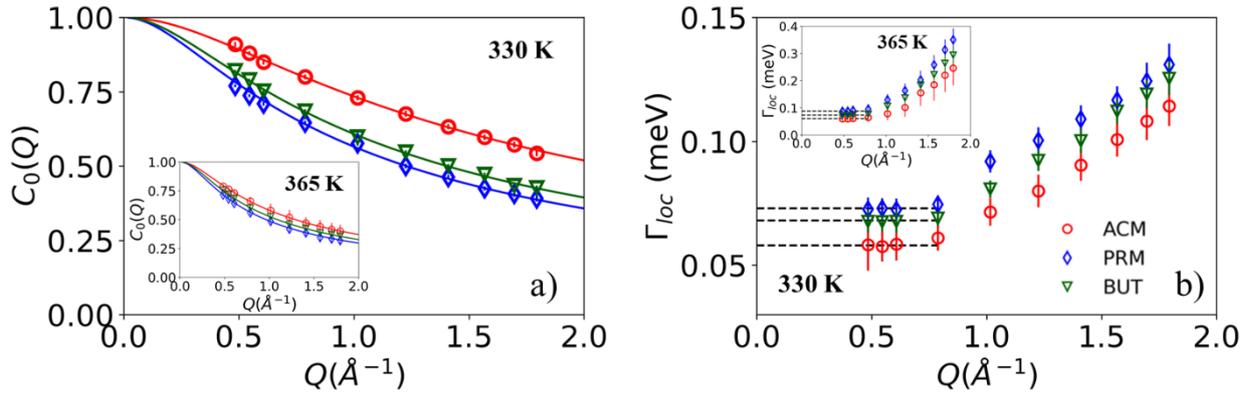

**Figure 5.** *(a) EISF, $C_0(Q)$, and (b) $\Gamma_{loc}$ verses $Q$ plot corresponding to localized motion describing the localized translation diffusion of hydrogen atoms for all three DESs at 330 K and 365 K (inset). The solid lines indicate the respective fits using Eq. 9 based on LTD model*

**Table II.** Dynamical Parameters obtained from fitting of EISF with Eq. (9).

| Amide | $\sigma_0$ (Å) | | $\tau$ (ps) | |
|---|---|---|---|---|
| | 330 K | 365 K | 330 K | 365 K |
| Acetamide | 0.54 (± 0.01) | 0.90 (± 0.01) | 10.34 (± 0.01) | 10.00 (± 0.01) |
| Propanamide | 0.94 (± 0.01) | 1.20 (± 0.01) | 8.22 (± 0.01) | 6.82 (± 0.02) |
| Butyramide | 0.82 (± 0.02) | 1.06 (± 0.02) | 8.82 (± 0.01) | 8.22 (± 0.03) |

#### 4.2.1 Delineating behavior of internal dynamics using MD simulations

Trajectories from the MD simulations were used to study the cage-jump diffusion and molecular flexibility of alkylamide in these DESs, to be able to delineate the individual contributions underlying the observed trend in QENS. The contributions of caged motions and jump dynamics are explored by modelling the mean squared displacement (MSD) of the alkylamide COM as a function of time. Meanwhile, the role of molecular flexibility is examined by identifying the range of conformational degrees of freedom accessible to the molecules. A comprehensive discussion of these findings is present in the subsequent subsections.

**Cage-Jump Diffusion**

As depicted in Figure 6, MD simulation trajectories of the COM for individual alkylamides clearly reveal the formation of cage-like clusters and instances of jump-like motion, a phenomenon well-aligned with the cage-jump diffusion model employed in the analysis of QENS spectra.

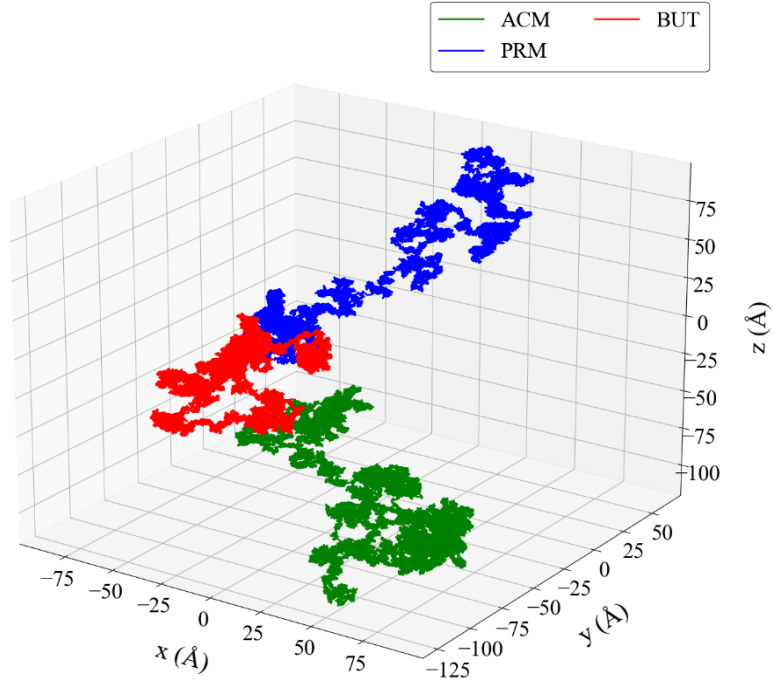

**Figure 6.** *COM trajectory of ACM (red), PRM (blue) and BUT (green) from MD simulations over 20 ns*

To further quantify the diffusive behavior of alkylamides in this system, the MSD was utilized as a key analytical tool. The MSD of the alkylamide COM was calculated from the simulation data collected over a total time interval of 20 ns and was determined using the following equation:

$$\langle \delta r^2(t) \rangle = \frac{1}{N} \sum_{i=1}^{N} \langle |r_i(t+t_0) - r_i(t_0)|^2 \rangle \qquad (10)$$

with $r_i(t)$ representing the position of the COM of the i$^{th}$ alkylamide molecule at time $t$ and $N$ being the total number of alkylamide molecules in the system. The averaging is done over different time origins, $t_0$, as indicated by angular brackets in the equation.

    The log-log plot showing the temporal variation of the MSD in Figure 7(a), distinctly highlights the presence of three different diffusion processes across various time scales, each reflecting unique molecular dynamics. At very short timescales (less than 0.1 ps), the MSD shows ballistic behavior, characterized by a $t^2$ dependance. In this regime, the molecules move inertially without significant interaction with their neighbors, essentially reflecting free, unimpeded motion. A closer analysis of the figure reveals that ACM exhibits more pronounced inertial movement than BUT, with a higher MSD in this ballistic region. This can be attributed to the fact that inertial motion is closely related to the mass of the molecules. As chain length increases—from acetamide to propanamide to butyramide—the molecules become heavier, leading to slower ballistic motion. The MSD behavior below 0.1 ps clearly reflects this trend, with acetamide showing the most rapid ballistic motion and butyramide the slowest.

As time progresses, the system transitions into a sub-diffusive regime, where the MSD evolves with a power-law dependence of $t^\alpha$, where $\alpha < 1$. This phase reflects the trapping of molecules inside transient cages formed by their neighboring molecules, resulting from intermolecular hydrogen bonding and complexation.[33] The duration and extent of this sub-diffusive regime provide critical insights into the strength and dynamics of these local interactions. Finally, as the cages relax and the molecules start exhibiting jumps from one cage to another, the MSD exhibits a linear dependence with time. This marks the onset of Fickian dynamics. Over longer time scales, as the number of cage-to-cage jumps increases, the motion of the molecules transitions into a diffusive phase, where the self-diffusion coefficient of the alkylamides COM can be determined using Einstein's relation.[73] This self-diffusion coefficient, calculated from the slope of the linear MSD region is equivalent to the jump diffusion coefficient obtained from QENS analysis. A similar trend in the self-diffusion coefficient as a function of chain length was observed, consistent with QENS data as shown in Table III, with the highest diffusion in acetamide and the lowest in butyramide.

The cage-jump diffusion model, previously used to interpret the QENS data was employed as a framework for describing the MSD behavior of the alkylamide COM. As outlined earlier, this model integrates two distinct dynamical modes: localized diffusion within transient cages and jump diffusion between these regions. Here we model the MSD data relevant to the timescales observable in the QENS instrument. QENS experimental measurements can typically discern variations in relaxation rates of the order of one-fifth of the instrument resolution (~17.5μeV) which roughly translates into a relaxation timescale of 4 to 5 times of its measurable time scale. As a result, the MSD is modeled in the range 1 -150 ps, ensuring it captures the relevant time scales for both localized and jump diffusion processes that are observed in the QENS instrument.

From the perspective of stochastic processes, the moments of the displacement can be obtained from self-intermediate scattering function (SISF), $I(Q,t)$, through differentiation. The $n^{th}$ moment of displacement is then expressed as:

$$\langle \delta r^{2n}(t) \rangle = (-1)^n \lim_{Q \to 0} \frac{\partial^{2n} I(Q,t)}{\partial Q^{2n}} \qquad (11)$$

The SISF contains comprehensive information, both temporal and geometrical aspect, about the different diffusion processes in the system and is related to incoherent scattering function, $S_{inc}(Q,E)$ by an inverse Fourier transform with respect to time. Therefore, it can be used to delineate the different diffusive components involved in the motion of the molecules in the system. Since localized diffusion within cages and jump diffusion events are independent processes, the effective SISF for the overall cage-jump diffusion process using the cage-jump diffusion model is obtained as the product of their individual contributions:

$$I(Q,t) = \exp\left[\frac{-(Qx_0)^2}{1+(Qx_0)^2}\frac{t}{\tau_j}\right] \langle \exp[-(Q\sigma)^2(1-\rho(t))] \rangle_\sigma \qquad (12)$$

The first term accounts for jump diffusion, while the second term describes the contribution from localized (caged) diffusion, modeled using the LTD model under the assumption of an isotropic Gaussian potential well.[72] Here, $x_0$ and $\tau_j$ represents the characteristic jump length and the average time interval between consecutive jumps, respectively. The symbol $\langle \rangle_\sigma$ denotes an average over the distribution of $\sigma$, the characteristic localization radius. The distribution of $\sigma$ is modeled using an exponential distribution, as previously described in the QENS analysis of internal dynamics. The rate of localized diffusion process, $\rho(t)$, follows an exponential relaxation function of the form $e^{-t/\tau_0}$, with $\tau_0$ the characteristic relaxation time. It follows that, from this model MSD can be derived as

$$\langle \delta r^2(t) \rangle = 6\left[\left(1 - e^{-t/\tau_0}\right) 2\sigma_0^2 + D_j t\right] + B \quad (13)$$

where $D_j = \frac{x_0^2}{\tau_j}$ is the jump diffusion coefficient, and B is a constant, which takes into consideration of the remaining contributions. Using Eq. (13) we modelled the obtained MSD with $\tau_0$, $D_j$, $\sigma_0$ and $B$ as fitting parameters in a time frame ranging from 3 to 150 ps. The model is found to capture the variation in MSD fairly well for all three systems as shown in Figure 7(b). The dynamical parameters obtained from the fit are tabulated in Table III.

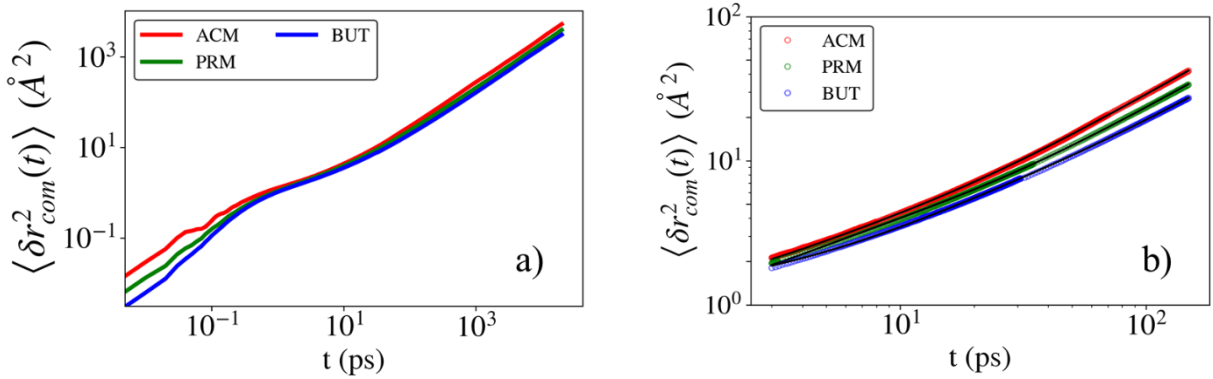

**Figure 7.** *a) Log-log scale plot showing the typical variation of MSD of alkylamide COM as function of time and the b) MSD fit using cage-jump diffusion model. The solid lines in black are the respective fits*

The parameters $\sigma_0$ and $\tau_0$ play a crucial role in understanding the cage dynamics within the system. Specifically, $\sigma_0$ represents the average value of the distribution function i.e. the cage radius, while $\tau_0$ characterizes the average timescale over which caged motion occurs. The data indicates that the cage radius, $\sigma_0$, remains relatively consistent across different chain lengths, with only minor variation. ACM exhibits a slightly smaller cage radius compared to PRM and BUT, which display almost identical radii. This suggests that the spatial confinement of molecules is not significantly influenced by the alkyl chain length. The relaxation dynamics of the caged motion, as quantified by the characteristic relaxation time $\tau_0$, reveal distinct differences across the alkylamide systems. BUT, with its longer alkyl chain, demonstrates significantly slower caged dynamics, evidenced by an over threefold increase in $\tau_0$ relative to PRM. This pronounced increase indicates that BUT molecules experience prolonged confinement within their local cages, likely due to the stronger intermolecular interactions, which impede the mobility of the molecules and slow down their localized motion. In contrast, ACM and PRM exhibit nearly identical relaxation timescales, with only a slight variation in $\tau_0$. The similar relaxation timescales suggest that the transition from ACM to PRM does not significantly alter their caged dynamics, as the timescale governing localized motion remains consistent between the two systems. However, the longer alkyl chain of BUT leads to a marked slowdown in caged dynamics, indicating that while the relaxation timescales for ACM and PRM are nearly identical, the increased chain length in BUT introduces greater spatial restriction and slower molecular relaxation within the cage.

In addition, the jump diffusion coefficient obtained from the fits also corroborates with the jump diffusion coefficients derived from the long-time limit slope of the MSD, calculated using Einstein's relation for diffusion. [73]

| Amide | Cage-Jump model | | | Linear regime |
|---|---|---|---|---|
| | $\sigma_0$ (Å) | $\tau_0$ (ps) | $D_j$ (x $10^{-6}$ cm$^2$/s) | $D_j$ (x $10^{-6}$ cm$^2$/s) |
| Acetamide (ACM) | 0.28 | 4.00 | 4.54 | 4.41 |
| Propanamide (PRM) | 0.35 | 5.00 | 3.58 | 3.26 |
| Butyramide (BUT) | 0.37 | 18.01 | 2.75 | 2.56 |

**Table III**: Dynamical Parameters obtained from fitting of MSD data from MD trajectories for the three systems at 365 K

**Molecular Flexibility**

The flexibility of alkylamides molecules is another major contributor to the localized dynamics observed in the QENS spectra. This flexibility of the molecules is explored through its conformational changes. For this, we principally consider two parameters – (i) $r_{CN}$, average distance between terminal carbon ($C_T$) and nitrogen (N) atoms from the molecular center and (ii) $\theta$, the angle subtended by vectors joining the molecular center to $C_T$ and the molecular center to N. Figure 8 depicts a schematic of different alkylamide molecules, illustrating the above discussed parameters with $r_{CN} = \frac{r_1 + r_2}{2}$, where $r_1$ and $r_2$ are the distances of terminal carbon and nitrogen atoms from molecules center.

The typical variation of $r_{CN}$ and $\theta$ with time for different alkylamides are presented in the supporting information (SI). Clearly, it is evident from these plots (Figure S2) that BUT explores a wide range of conformations and ACM explores a very narrow range. The conformational landscape explored by these molecules can be quantified by calculating the probability distribution, $p(r_{CN}, \theta)$, that gives the probability of molecule being in the conformation with the distance $r_{CN}$ with an angular bend of $\theta$.

To assess the flexibility of alkylamides, a contour plot was employed to capture the behavior of $p(r_{CN}, \theta)$. The x-axis of the contour represents $\theta$, and y-axis, represents $r_{CN}$. The contour plot provides a comprehensive overview of both angular and spatial flexibility, illustrating how the angle subtended varies with their spatial distance and highlighting how chain length influences the internal flexibility and conformational behavior of alkylamides. The contour maps of ACM, PRM and BUT are presented in Figure 8 with color gradient, ranging from light pink to deep purple, representing the magnitude of the probability distribution. For

ACM, a single sharp region is observed, centered at 1.38 Å with an angular orientation of 122°. The sharpness of this peak at a single value suggests that the internal dynamics within ACM are highly restricted, indicating an almost rigid molecular structure, which is consistent with its short alkyl chain. The concentrated region indicates limited spatial freedom, suggesting that ACM's internal flexibility is minimal.

Contrastingly PRM, displays two distinct regions positioned at 1.75 Å and 1.9 Å, with angular orientations of 120° and 160°, respectively. The appearance of two regions in the contour plot reflects increased conformational diversity compared to ACM. The broader angular spread between the regions indicates greater flexibility, implying that the longer alkyl chain of PRM allows for more varied spatial configurations. These variations in spatial positions are indicative of the molecule's ability to adopt multiple conformational states, contributing to its overall flexibility in PRM relative to ACM. The favorable conformations for both ACM and PRM are shown in Figure S3. In the case of ACM, the most stable conformation is observed when the orientation between the terminal carbon and nitrogen atoms forms an angle of 120°. This single, well-defined conformation reflects the restricted spatial flexibility of ACM due to its shorter alkyl chain, limiting the molecule to a narrow range of internal configurations. In contrast, PRM exhibits not only this 120° conformation but also an additional favorable state at an angular orientation of 160°. The appearance of this second conformational state highlights the influence of PRM's longer alkyl chain, which provides greater flexibility and allows the molecule to adopt a wider range of spatial configurations.

In case BUT, three distinct regions are observed at angular orientations of 90°, 127°, and 170°, all positioned at 2.5 Å. The larger angular displacement observed in BUT can be attributed to its longer alkyl chain, which includes an additional -$CH_2$- group, in comparison to its counterparts. The presence of three distinct angular regions indicates greater internal flexibility, allowing BUT to explore a wider range of conformations. The broad angular distribution, spanning from 90° to 170°, highlights BUT's ability to adopt multiple distinct angular configurations. This conformational diversity is reflected in the three favorable states shown in the Figure S3. The ability of BUT to adopt both extended and compact configurations underscores the increased conformational freedom associated with longer alkyl chains.

These observations collectively indicate that as the alkyl chain length increases, the molecule exhibits a broader range of conformational states which manifest as increased internal flexibility. This trend is evident from the restricted dynamics of ACM, the intermediate flexibility of PRM, and the highly flexible nature of BUT, suggesting that alkyl chain length is a key factor in determining the flexibility and internal dynamics of alkylamides.

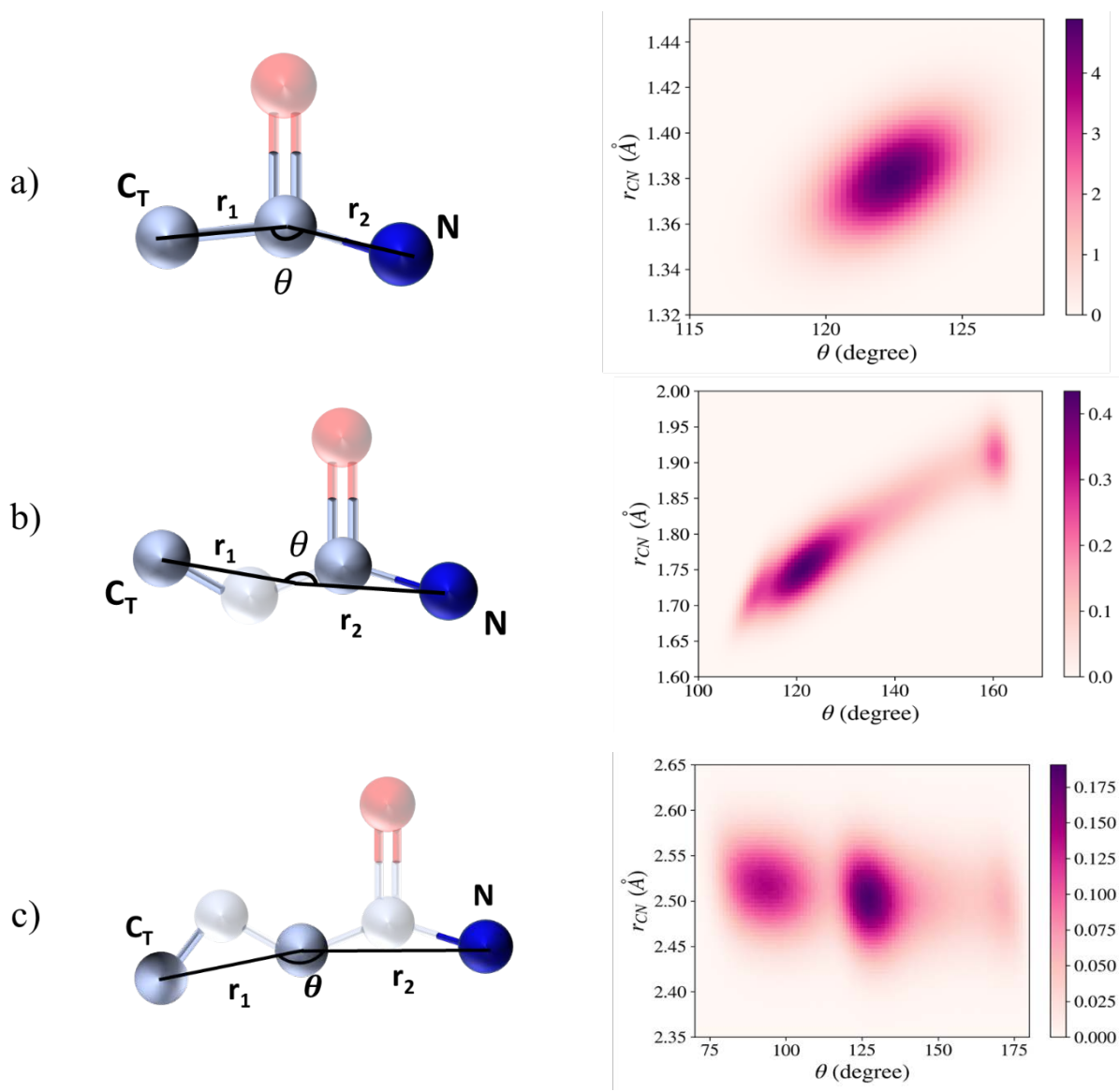

**Figure 8**. (Left) *Schematic representation of alkylamide molecules showing the distance of terminal carbon ($C_T$) and nitrogen atoms (N) atoms from the molecular center and the angle subtended ($\theta$) between them (Right) Contour plots illustrating the variation of angular distribution ($\theta$) with spatial distribution ($r_{CN}$) between the terminal atoms (carbon of $CH_3$ and nitrogen of $NH_2$) for a) ACM b) PRM and c) BUT respectively.*

To consolidate these findings, it is clear that two primary factors—caged motion and molecular flexibility—play crucial roles in shaping the internal dynamics of alkylamides within DESs. Although the caged dynamics of ACM and PRM are nearly identical, the longer alkyl chain of PRM increases its molecular flexibility, allowing for greater internal motion. This enhanced flexibility contributes to faster internal dynamics in PRM compared to ACM, despite the minimal differences in their caged relaxation timescales. The additional chain length in PRM permits more pronounced internal degrees of freedom, facilitating quicker molecular rearrangements and diffusion on short timescales. When comparing PRM and BUT, the longer alkyl chain of BUT confers even greater molecular flexibility. However, this increased flexibility is counterbalanced by significantly slower caged dynamics, as indicated by the much longer relaxation time $\tau_0$. The restricted molecular mobility within the cage for

BUT dominates the overall dynamics, causing it to exhibit slower internal motion than PRM. Thus, despite BUT's greater flexibility, the slower caged motion results in PRM having faster overall internal dynamics as observed in QENS experiments.

Thus, PRM emerges as the fastest of the three alkylamides in terms of localized motion. Its moderate flexibility, combined with relatively fast caged dynamics, enables it to achieve more rapid molecular rearrangements compared to both ACM, with its more rigid structure, and BUT, which, despite its flexibility, is hindered by its slower caged motion.

## 5. Conclusion

This study, combining data from QENS and MD simulations, elucidates the impact of alkyl chain length on the nanoscopic diffusive dynamics of alkylamides in DESs synthesised from different alkylamides and lithium perchlorate salt. Our analysis reveals that the overall dynamics of the alkylamides are governed by two distinct types of motions: (i) long-range jump diffusion of the alkylamide COM and (ii) localized diffusion within confined volumes. Both types of motion are notably influenced by the alkyl chain length. A pronounced slowdown in jump dynamics is observed with increasing alkyl chain length as evidenced by a systematic decrease in the jump diffusion coefficient of the alkylamide COM. This deceleration is primarily attributed to two factors: (i) the increased size and mass associated with longer alkyl chains, and (ii) the enhanced complexation between the alkylamides and lithium ions as the chain-length increases.

In contrast, localized diffusion of alkylamides does not exhibit a simple monotonic dependence on chain length. Conformational analysis reveals that longer alkyl chains enhance the molecular flexibility of alkylamides, enabling them to explore a broader range of conformational states. For instance, acetamide with its shorter chain, predominantly exhibits a single stable conformation, whereas butyramide accesses multiple conformational states. This increased flexibility in butyramide facilitates a broader range of internal configurations, potentially leading to increased localized dynamics.

Finally, experimental results reveal that propanamide exhibits the highest localized mobility, followed by butyramide and acetamide, which highlights the role of additional factors beyond molecular flexibility in shaping the localized dynamics within DESs. MD simulations show that acetamide and propanamide exhibit relatively fast caged dynamics, whereas butyramide experiences the slowest. The interplay between these factors results in propanamide achieving the fastest localized dynamics. Conversely, despite its significant molecular flexibility, butyramides localized dynamics are hindered by its slower caged dynamics, resulting in reduced overall localized motion compared to propanamide.

In summary, the interplay between molecular flexibility and caged dynamics plays a crucial role in governing the localized motion of alkylamides. While propanamide benefits from the dual effects of enhanced flexibility and accelerated caged dynamics, butyramides greater flexibility is offset by slower caged dynamics, resulting in slower overall motion compared to propanamide.

These findings underscore the complex interplay between alkyl chain length and the nanoscale dynamics in DESs. They demonstrate that by modulating the molecular structure of individual components, the dynamic properties of DESs can be effectively tailored, enabling the design of solvents with tailored dynamic and physicochemical properties for specific industrial and scientific applications.

**Supporting Information**

Refer to the supplementary material for data on the intermolecular probability distribution function between lithium ions and oxygen sites of alkylamides as a function of chain length, the probability distribution illustrating the temporal variation of spatial and angular conformations of alkylamides and information on the force field parameters employed in the MD simulations.

**Acknowledgements**

We acknowledge the Computer Division, Bhabha Atomic Research Centre, India, for their generous support in providing computational time on the Atulya supercomputers.

**Conflict of Interest**

The authors declare no conflict of interest.

**Data Availability Statement**

The data that supports the findings of this study are available from the corresponding author upon reasonable request.

**Keywords**

Molecular Dynamic simulations, Neutron Scattering, Diffusion, Deep Eutectic Solvents, electrolyte DESs